# Genetic Code: The unity of the stereochemical determinism and pure chance

Miloje M. Rakočević

*Faculty of Science, University of Niš (now retired, on the Address: Milutina Milankovica 118, 11070 Belgrade, Serbia  (e-mail: m.m.r@eunet.yu; or: milemirkov@nadlanu.com; www.rakocevcode.rs)*

**Abstract**

It is presented that the positions of amino acids within Genetic Code Table follow from strict their physical and chemical properties as well as from a pure formal determination by the Golden mean.

In this short paper the main result of our previous work (Rakočević, 2009) is presented in a new light; in the light of Crick's two hypotheses, put forward more than forty years ago (Crick, 1968). Crick asked then: Is the genetic code the result of strict stereochemical determinism, or of pure chance? Since those times many different answers have replaced one another. However, we can say that the first hypothesis has prevailed, though occasionally there still appears the second one as a possibility. [Crick himself left this question as an open question for further research.]

Our main result from the previous work shows that there follows from the Genetic Code Table (GCT) (Table 1) and the corresponding binary-code-tree (Figure 1 in relation to Table 2) a strict classification of amino acids (AAs) into five purely chemical classes (Table 3). And that is the said strict chemical determinism; moreover, that determinism is *stereochemical* precisely because the organization of amino acids into pairs must take into account the fact that the nature of the canonical twenty amino acids is such that they necessarily exist in the form of four stereochemical types (Popov, 1989; Rakočević and Jokić, 1996);[1] glycine within the glycinic type, proline within the prolinic, valine and isoleucine within the valinic, and the remaining 16 amino acids within the alaninic stereochemical type – the 12 aliphatic and four aromatic amino acids. [*Remark* 1: "The same four stereochemical types, two and two (G, A and V, P) ... also come from the amino acid constitution structures, in the following manner. The side chain of glycine (-H) comes from the shortest possible hydrogen chain (H-H) and none of 19 other amino acids has a hydrogen chain of this kind. The side chain of alanine ($-CH_3$), or, in relation to glycine, ($-CH_2-H$), follows from the shortest possible non-cyclic hydrocarbon chain ($CH_4$), and still 15 amino acids have the alanine-analogue side chain in the form ($-CH_2-R$). The side chain of valine ($H_3C-CH-CH_3$) follows from the shortest possible cyclic hydrocarbon, from cyclopropane, with a permanent openness and with a linkage to the "head" of the amino acid through only one vertex of the cyclopropane "triangle". ... The proline type (only with proline) follows from the same source (cyclopropane), but with a permanent non-openness and with a linkage to the "head" through two vertices of the cyclopropane "triangle"" (Rakočević and Jokić, 1996, p. 345)[2].]

---

[1] It is simply amazing that, in genetic code research, no one else but us has referred to this work by Popov.

Besides, one step before the system "5 x 4" there is a system ["(2x7)"+"(1x6)"] (Survey 1) with seven "golden" AAs, the order of which follows from the strict golden mean determination; a determination which, in itself, is not physical, or chemical, but of a purely formal nature. It follows from this, as direct evidence, that it is the question of pure chance[3] whether or not (in the prebiotic states of matter) the conditions can be created for the appearance of an adequate molecular aggregation; adequate with respect to golden mean determination[4].

But, let's look once again at what follows from what. First we look at Survey 1 in relation to Table 2. Exactly seven AAs are on the positions of the golden mean, $\phi^n$ (n = 0, 1, 2, 3, ... , 9). They are G, Q, T, P, S, L, F. Across from them there are seven of their pairing "partners" (within the pairs), as the complements, and below are the six remaining AAs, the three pairs of the noncomplements: D-E, K-R, H-W.

---

[2] Giving to all this the presence of the first possible branching (coming from isobutane) in leucine and isoleucine, one can say that the genetic code is the result of the first possible cases: the first possible openness, ciclicity, half-ciclicity and branching. If so, if these extreme cases represent the condition for the origin of the code, then the possibility of origin of the life is significantly reduced anywhere in the universe.

[3] In this moment I bear in mind the famous book written by Jacques Monod (Chance and necessity, Vintage books, ISBN 0-394-71825-9) in which on the page V is cited the great Democritus of Abdera, living in the 4th century BC: "Everything existing in the Universe is the fruit of chance and of necessity".

[4] This question has also an exobiological aspect. Namely, if the determination by golden mean (the correspondence with the best possible symmetry, proportion and harmony) is the condition for the origin of the code, then it significantly reduces the possibility of origin of the life anywhere in the universe.

[And it makes sense that aliphatic AAs come first (D-E/K-R), followed by aromatic ones (H-W), all of them in the order determined by the size of the molecules, i.e. by the number of atoms within the molecule (side chain).]

In order to see all this, one must previously establish the order of codons in the GCT. Immediately it is obvious that the codon UUU is the zeroth one, and the codon CUG the seventh (8$^{th}$ in the series of codons). After that, the problem is whether AUU or UCU is the eighth codon (9$^{th}$ in the series)? In other words, what type of the base precedes in the chemical hierarchy: purine-pyrimidine-any (Pu-Py-X) or pyrimidine-pyrimidine-any (Py-Py-X)? Over ten years ago, I showed (Rakočević, 1998) that the determination by the golden mean coincides with the chemical order in which the Py-Py-X codons must precede the Pu-Py-X ones (as it is presented in Figure 1 in correspondence to Table 1). [*Remark* 2: The presented order of codons is based on the codon octets, but for a better viewing how the system, presented in Table 3, follows from GCT, it is also necessarily to have an order based on the codon quartets, such as it is presented in Table 4 (Negadi, 2009). Also to have a very symmetric form of GCT, valid for mitochondrial genetic code (Dragovich & Dragovich, 2006, 2007a, 2007b]

So, from the codon order follows the order of AAs, an order where seven of them are found at the positions of the golden mean. But, if the sequence of the seven AAs was turned through 180 degrees, and then the arrangement was made by amino acid molecule size, first glycine with one atom in the side chain, serine with 5, threonine with 8 atoms (the pair T 08 / M 11 must be before the pair of P 08 / I 13) etc., we would get the order as in Table 3: G, S, T, P, Q, L, F. We actually got a strict system that we in the previous paper (Rakočević, 2009) called CIPS (Cyclic Invariant Periodic System)[5].

The following analysis shows that the order of five of the amino acid classes makes sense. On the position "1" came AAs of non-alaninic type, which are aliphatic by origin: G-P and V-I; followed (on the position "2") by aliphatic AAs of the alaninic type (A-L and K-R), two of which are amine derivatives (K-R), with a lower degree of polarity (nitrogen is less polar then oxygen!)

From pure chemical reasons it makes sense to say that these two classes (light tones in Table 3) belong to *a primary superclass*, with original aliphatic AAs (and/or derivatives of lower level), whereas the three remaining classes (dark tones) to *a secondary superclass*, with the derivatives of a higher level.

---

[5] Cyclicity and periodicity through the positions of two and two amino acids – up/down – in relation to middle chalcogene AAs, on the position "3".

On the position "3" within the system, presented in Table 3, came chalcogene AAs (S, T & C, M); followed (on the position "4") by two double acidic AAs with two their amide derivatives (D, E & N, Q); finally came four aromatic AAs (F,Y & H, W) on the position "5".

The splitting into two superclasses exists in a strict correspondence to the splitting into two cathalitic directed classes (amino acids handled by class I or by class II of aminoacyl-tRNA synthetases), as it is shown in Table 5, in relation to Survey 2. From the said correspondence can follow *a prediction* for further researches: the demonstrated crossing of two classes and two superclasses must be reflected (in some way?) in the protein structures and functions.

As it is immediately obvious from CIPS[6], the nature of the genetic code again points out the validity of Aristotle sentence[7], and, on the other hand, necessarily leads to the conclusion that both Crick's hypotheses about the origin of the genetic code, *mutatis mutandis*, are valid.

---

[6] Which system is the "fruit" of a unity of strict stereochemical determinism and pure chance, a unity of physical–chemical characteristics and pure formalism (cf. footnote 3).

[7] "The existence of such a harmonic structure with unity of a determination with physical–chemical characteristics and atom and nucleon number ... appealed to Aristotle and to his idea of unity of form and essence" (Rakočević, 2004, p. 233).

| 1st lett. | 2nd letter | | | | | | | | 3rd lett. |
|---|---|---|---|---|---|---|---|---|---|
| | U | | C | | A | | G | | |
| U | 00. UUU | F | 08. UCU | S | 32. UAU | Y | 40. UGU | C | U |
| | 01. UUC | | 09. UCC | | 33. UAC | | 41. UGC | | C |
| | 02. UUA | L | 10. UCA | | 34. UAA | CT | 42. UGA | CT | A |
| | 03. UUG | | 11. UCG | | 35. UAG | | 43. UGG | W | G |
| C | 04. CUU | L | 12. CCU | P | 36. CAU | H | 44. CGU | R | U |
| | 05. CUC | | 13. CCC | | 37. CAC | | 45. CGC | | C |
| | 06. CUA | | 14. CCA | | 38. CAA | Q | 46. CGA | | A |
| | 07. CUG | | 15. CCG | | 39. CAG | | 47. CGG | | G |
| A | 16. AUU | I | 24. ACU | T | 48. AAU | N | 56. AGU | S | U |
| | 17. AUC | | 25. ACC | | 49. AAC | | 57. AGC | | C |
| | 18. AUA | | 26. ACA | | 50. AAA | K | 58. AGA | R | A |
| | 19. AUG | M | 27. ACG | | 51. AAG | | 59. AGG | | G |
| G | 20. GUU | V | 28. GCU | A | 52. GAU | D | 60. GGU | G | U |
| | 21. GUC | | 29. GCC | | 53. GAC | | 61. GGC | | C |
| | 22. GUA | | 30. GCA | | 54. GAA | E | 62. GGA | | A |
| | 23. GUG | | 31. GCG | | 55. GAG | | 63. GGG | | G |

**Table 1.** The Table of the standard genetic code. Ordinal number of codons after the order-key: YYN, RYN, YRN, RRN, in correspondence with the hierarchy on the binary-code tree in Figure 1. [One-letter abbreviations: Y from pYrimidine; R from puRine; and N from aNy (of bases).]

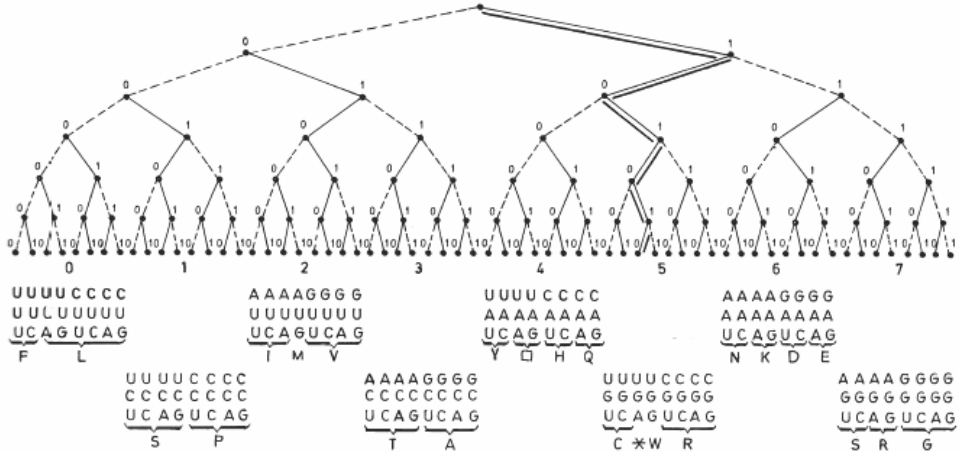

**Figure. 1**. Genetic code as a binary-code tree. The full lines: the routes of the greater changes, from 0 to 1 and vice versa; the dotted lines: the routes of the less changes, from 0 to 0, as well as from 1 to 1 going from a higher into a lower level. The double full line: the route of the maximum possible changes: from 0 to 1 and vice versa in any step (the route corresponding to the 'Golden mean route' on the Farey tree). The codon order after the rules given by R. Swanson (Swanson, 1984, Figure 1): 1 for purine and 0 for pyrimidine; 1 for three and 0 for two hydrogen bonds.

| $\varphi^0$ | $\varphi^1$ | $\varphi^2$ | $\varphi^3$ | $\varphi^4$ | $\varphi^{5-7}$ | $\varphi^8$ | $\varphi^9$ |
|---|---|---|---|---|---|---|---|
| G | Q | T | P | S | L | L | F |
| 63 | 39-38 | 25-24 | 15-14 | 10-09 | 06-02 | 02-01 | 01-00 |
| 63 | 38.94 | 24.06 | 14.87 | 9.19 | 5.68 – 2.17 | 1.34 | 0.83 |

**Table 2.** The amino acids in Golden mean power positions within the sequence 0–63 on the binary-code tree in Fig. 1. First row: Golden mean powers within first 'cycle' in module 9. Second row: amino acids in the positions marked in third row, taken from the binary-code tree in Fig. 1. Fourth row: the values of the Golden mean powers within the interval 0–63. The calculations: 0.618033 x 63 = 38.94; 0.618033 x 0.618033 x 63 = 24.06 etc.

| | | | | | | |
|---|---|---|---|---|---|---|
| 5 | 073 | F | 14 | 15 | Y | 079 |
| 4 | 235 | L | 13 | 04 | A | 172 |
| 3 | 087 | Q | 11 | 08 | N | 085 |
| 2 | 160 | P | 08 | 13 | I | 121 |
| 1 | 168 | T | 08 | 11 | M | 043 |
| 1 | 243 | S | 05 | 05 | C | 081 |
| 2 | 184 | G | 01 | 10 | V | 168 |
| 3 | 087 | D | 07 | 10 | E | 093 |
| 4 | 091 | K | 15 | 17 | R | 265 |
| 5 | 081 | H | 11 | 18 | W | 044 |

**Table 3.** The Cyclic Invariant Periodic System (CIPS) of canonical AAs. At the outer side, left and right, it is designated the number of atoms within coding codons; more exactly, in the Py-Pu bases (U = 12, C = 13, A = 15 and G = 16); at the inner side – the atom number within amino acid side chains. In the middle position there are chalcogene AAs (S, T & C, M); follow - in next „cycle" - the AAs of non-alaninic stereochemical types (G, P & V, I), then two double acidic AAs with two their amide derivatives (D, E & N, Q), the two original aliphatic AAs with two amine derivatives (A, L & K, R); and, finely, four aromatic AAs (F,Y & H, W) – two up and two down. The said five classes belong to two superclasses: primary superclass in light areas and secondary superclass in dark areas. Notice that each amino acid position in this CIPS is strictly determined and none can be changed. [For details see the text; for arithmetical regularities, atom and nucleon number balances within this CIP system see in our previous paper (Rakočević, 2009).]

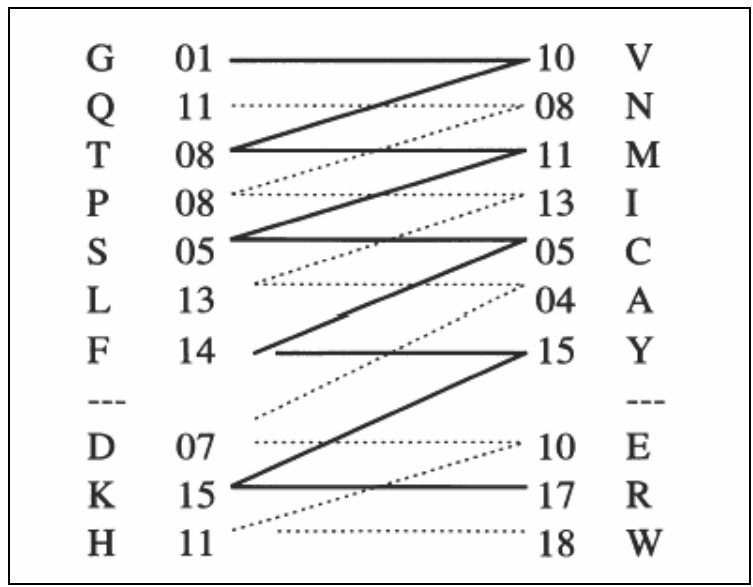

**Survey 1.** Atom number balance directed by Golden mean on the binary-code tree (Scheme 2 in Rakočević, 1998; Table 3 in Rakočević, 2009). First seven amino acids on the left are 'golden' amino acids with 60 atoms within side chains; on the right are their complements with [60 + (1 x 6)] = 66 atoms; below are three amino acid pairs as non-complements with [66 + (2 x 6)] = 2 x 39 = 78 atoms; Notice that within aliphatic non-complements there are 39+10, whereas within aromatics (H & W) 39-10 of atoms; in the other words: (8 x 6) +1 and (5 x 6) -1, respectively. On the first zigzag (full) line there is 102-1 whereas on the second (dotted) line 102+1 atoms. Arithmetic mean for both: 102±1. Notice also that the arrangement-ordering of "golden" AAs is the same as in Table 2.

```
U-U-X   U-C-X   C-U-X   C-C-X
U-A-X   U-G-X   C-A-X   C-G-X
A-U-X   A-C-X   G-U-X   G-C-X
A-A-X   A-G-X   G-A-X   G-G-X
```

**Table 4 .** The codon order in Genetic Code Table, based on the codon quartets in first and second half of the Table. [After Table 1 in (Negadi, 2009), here simplified and generalized.]

| 28 | 09 | G P | (1) | 23 | V I | 53 | 81 |
|---|---|---|---|---|---|---|---|
|    | 19 | A K | (2) | 30 | L R |    |    |
|    |    |     |     |    |     |    |    |
|    | 13 | S T | (3) | 16 | C M |    |    |
| 53 | 15 | D N | (4) | 21 | E Q | 70 | 123 |
|    | 25 | F H | (5) | 33 | Y W |    |    |
| 81 |    |     |     |    |     | 123 | 204 |

**Table 5.** The arrangement in accordance to the principle: "a little" and "full" in relation to "small" and "large". So, on the left there are AAs (81 atoms) from the left side of Survey 2 (class II, with smaller molecules within the pairs); and on the right there are AAs (123 atoms) from the right side of Survey 2 (class I, with larger molecules within the pairs). At the same time very up there are AAs from primary superclass (81 atoms), just aliphatic and nonpolar (A,V, L, I) and "a little" polar (G, P, K, R) (hydrogen and nitrogen are less polar then oxygen!); in the other hand, except aromatic and sulfur AAs, down are AAs from secondary superclass (the row with 123 atoms), also aliphatic, but "full" polar. (This Table is the same as Table D.2 in Rakočević, 2009.)

| | | | | | |
|---|---|---|---|---|---|
| 1 | G 01 | | | 10 V | (1 x 11)±0 |
| 3 | S 05<br>T 08 | 14 | 26 | 05 C<br>11 M | 1 x 29 |
| 1<br>2 | P 08<br>A 04 | 12 | 26 | 13 I<br>13 L | 2 x 19 |
| 4 | D 07<br>N 08 | 30 | 38 | 10 E<br>11 Q | 2 x 18 |
| 2 | K 15 | | | 17 R | (3 x 11) -1 |
| 5 | H 11<br>F 14 | 25 | 33 | 18 W<br>15 Y | 2 x 29 |
| | 81 | | | 123 | |
| | (102 ± 1) | | | | |

**Survey 2.** Two amino acid classes generated through the influence of two catalysts (Scheme 5 in Rakočević, 1998 and Table D.1 in Rakočević, 2009). On the left the smaller molecules of AAs, handled by class II of aminoacyl-tRNA synthetases, whereas on the right the larger molecules of AAs, handled by class I of synthetases. On the full line there are 102+1 and on the dotted one 102-1 of atoms. The distinct arrangement of five classes (and two superclasses) from CIPS (Table 3) is self-evident. By this one must notice the existence of a crossing valid only for the primary superclass; the crossing in relation to original system in Table 3. (Why P together with A, L, I one can read in Remark 1.)

**Acknowledgement**

I am grateful to Branko Dragovich and Vladimir Ajdačić for helpful, stimulating discussion.